\documentstyle[eqsecnum,aps]{revtex}
\draft
\begin{document}
\newcommand{\be}{\begin{equation}}
\newcommand{\ee}{\end{equation}}
\newcommand{\bea}{\begin{eqnarray}}
\newcommand{\eea}{\end{eqnarray}}

\title{\Huge Symmetry Reduction and Semiclassical Analysis of
Axially Symmetric Systems}
\author{Santanu Pal\thanks{email: santanu@veccal.ernet.in}}
\address{Variable Energy Cyclotron Centre \\
1/AF Bidhan Nagar \\
Calcutta 700 064, India}
\author{Debabrata Biswas\thanks{email: biswas@kaos.nbi.dk}
\thanks{Permanent address : Theoretical Physics Division,
Bhabha Atomic Research Centre, Bombay 400~085}}
\address{Center for Chaos and Turbulence Studies,
Niels Bohr Institute \\
Blegdamsvej 17,
Copenhagen $\O$, Denmark}
\maketitle
\begin{abstract}
We derive a semiclassical trace formula for a symmetry reduced
part of the spectrum in axially symmetric systems. The classical
orbits that contribute are closed in $(\rho,z,p_\rho,p_z)$
and have $p_\varphi = m\hbar$ where $m$ is the azimuthal
quantum number. For $m\neq 0$, these orbits vary with
energy and almost never lie on periodic trajectories in the
full phase space in contrast to the case of discrete
symmetries. The transition from $m=0$ to $m > 0$ is however
not dramatic as our numerical results indicate, suggesting
that contributing orbits occur in topologically equivalent
families within which $p_\varphi$ varies smoothly.

\end{abstract}

\section{Introduction}
\label{sec:Intro}

\par Modern semiclassical theories deal with the duality
of the quantum energy eigenspectrum and the classical
spectrum of periodic orbit lengths and stabilities \cite{MCG}.
For integrable systems, the Poisson summation formula
provides this connection and is equivalent to the EBK
quantization scheme while for generic quantum systems,
trace formulas deal with this duality and provide the only
connection between quantum and classical mechanics.
As an example, for hyperbolic systems the Gutzwiller
trace formula expresses the density of quantum energy
eigenvalues, $d(E) = \sum_n \delta(E-E_n)$ as :

\be
d(E)  =  d_{av}(E) +
{1\over \pi \hbar} \sum_p \sum_{r=1}^\infty {T_p\over
\sqrt{\left|\det(M_p^r - I)\right|} } \cos \left(rS_p -
\pi \sigma_p r/2\right)
 \label{eq:GT}
\ee

\noindent
where the summation over $p$ extends over all primitive
periodic orbits, $T_p$ is the corresponding time period,
$S_p$ the action, $\sigma_p$ the Maslov index associated
with its invariant manifolds and $M_p$ the stability
matrix arising from a linearization of the transverse
flow. For systems where periodic orbits occur in families,
appropriate modifications need to be made
\cite{CL1,CL2}, though these may be system
specific \cite{PJ}.

\par Quite often, the systems we encounter in nuclear, atomic
or molecular physics have symmetries and one might be
interested in the spectrum of a particular symmetry
class.  For example, billiard systems with reflection
symmetry about the $x$ and $y$ axis have four classes
of wavefunctions with a choice of Dirichlet or Neumann
boundary condition on the symmetry lines.
In these as well as in other
systems with discrete symmetries, it is possible to
construct the symmetry reduced Green's function and
evaluate its trace \cite{Rob1}. While the orbits that
contribute
are periodic in the reduced phase-space,
they are not necessarily periodic in the full phase
space. The end-points however are related by symmetry.
It turns out nevertheless that when allowed to
evolve in time, these orbits are eventually periodic
in the full phase space. Thus trajectories
contributing to any given symmetry class of the spectrum
are either full periodic orbits or parts of these
whenever the symmetries present are discrete in
nature \cite{Rob1,CE}.

\par The presence of continuous symmetries
implies the existence of constants of motion.
For example in systems with axial  symmetry, the
$z$-component of the angular momentum is conserved
and quantum mechanically one may be interested in
the spectrum of a particular $m$-subspace. A way to
deal with this problem semiclassically is to work
with an effective two-dimensional
potential $V(\rho,z) + \hbar^2 m^2/(2M\rho^2)$ where $M$ is the
mass of the particle and $m$ is the azimuthal quantum
number. Here $V(\rho,z)$ is the actual potential
in which the dynamics is executed
and the  additional
term $\hbar^2m^2/(2M\rho^2)$ arises from the conservation of
$l_z$ \cite{cylindrical,WF}. The semi-classical
trace formula then takes the form in Eq.~(\ref{eq:GT})
provided periodic orbits are isolated and unstable.

\par
An alternative procedure is to work with the dynamics
in the full phase-space and it is desirable to
understand the nature of the trace formula and
the kind of classical orbits that contribute.
We shall devote ourselves to this question through
the rest of this paper and our results are in
certain sense very different
from those for discrete
symmetry. The orbits that contribute necessarily have
an angular momentum $l_z = m\hbar$ and must close
in $\rho, p_\rho$ and $z, p_z$. Thus they are
not necessarily periodic since they need not close in
$\varphi$. Unlike discrete symmetries however, orbits
that are not periodic are generally not parts of periodic
orbits.
We support these results with extensive numerical
evidence.

\par On completion of this work, we came
across a paper by Creagh \cite{Creagh} that deals
with this  problem  for arbitrary
Abelian symmetries and rotational symmetry.
Our approaches are
different however. Creagh \cite{Creagh} takes
recourse to group-theoretical methods in order to obtain
the symmetry  reduced Green's function
and then uses a semi-classical approximation that ultimately
involves classical trajectories possessing quantized
values of these additional constants of motion. Our
approach is based on the fact that for axially symmetric
systems, the eigenfunction,
$\psi(\rho,\varphi,z) = (2\pi)^{-1/2}e^{im\varphi} \phi(\rho,z)$
and we use this in the full Green's function to
obtain the density of a given $m$-subspace. The
restriction of constants of motion to their quantized
values arises from a stationary phase condition
in our case as also the fact that the trajectory
must close in $\rho, p_\rho$ and $z, p_z$.
Thus, even though the final results are identical, the
derivation in this paper provides an alternate viewpoint that
is simple and transparent to those not familiar with
the intricacies of extended phase space.

\par Apart from a re-derivation of the trace formula and
its numerical verification, this work emphasises the
fact that orbits contributing to a symmetry reduced
part of the spectrum are in general not eventually periodic
in the full phase space whenever the symmetry is
continuous. This is an important departure from
the case of discrete symmetry. We also show that
contributing orbits occur in topologically equivalent
families within which $p_\varphi$ varies smoothly.
For successive $m$ therefore, distinct orbits from the same
family contribute and this is evident from the
shift of peaks in the power
spectrum of the quantum density.

\par The organization of the paper is as follows. In
section \ref{sec:Formalism}, we provide a derivation of
the trace formula. Section \ref{sec:Length} deals with
the inverse problem that motivate
our numerical results which we present in
section \ref{sec:Numerical}. Our conclusions are
summarized in section \ref{sec:Conclusions}.


\section{Formalism}
\label{sec:Formalism}

Let  us  consider  an  axially symmetric three dimensional system
described by the coordinates $(\varphi,\eta)$ where $\varphi$
is  the  azimuthal  angle  about  the  axis   of   symmetry   and
$\eta=(\eta_2,\eta_3)$  represents the other two components in an
orthogonal coordinate system.

The  basic  definition  of  the  full  Green's function of such a
system can be written as ,
\be
 G(\varphi''\eta'',\varphi'\eta';E)
 =\sum_n\sum_m\frac{\psi_n^{(m)}(\varphi''\eta'')
\psi_n^{(m)^{\dag}}(\varphi'\eta')}{E-E_{n,m}} \label{eq:bas}
\ee
where
\be
 \psi_n^{(m)}(\varphi,\eta)=\frac{\exp(im\varphi)}{\sqrt{2\pi}}
    \phi_n^{(m)}(\eta) \label{eq:wave}
\ee
\noindent
In the above, both $\psi$ and $\phi$ are taken to be normalised and $m$
is the azimuthal quantum number. Thus :

\be
G(\varphi''\eta'',\varphi'\eta' ;E)
=\frac{1}{2\pi}\sum_m \exp[im(\varphi''-\varphi')]\sum_n
\frac{\phi_n^{(m)}
(\eta'')\phi_n^{(m)^{\dag}}(\eta')}{E-E_{n,m}} \label{eq:man1}
\ee

\noindent
Multiplying  both sides by $\exp(-i\mu\varphi'')$ and integrating
over  $\varphi''$, one obtains,

\be
\sum_n\frac{\phi_n^{(\mu)}(\eta'')\phi_n^{(\mu)^{\dag}}
(\eta')}{E-E_{n,\mu}}
=\int_0^{2\pi}
G(\varphi''\eta'',\varphi'\eta';E)
  \exp[-i\mu(\varphi''-\varphi')]d\varphi'' \label{eq:man2}
\ee

\noindent
where \{$E_{n,\mu}$\} is the subset of eigenvalues
for which the azimuthal quantum number equals $\mu$.
Further integration over $\varphi'$ on both sides gives,

\be
\sum_n\frac{\phi_n^{(\mu)}(\eta'')\phi_n^{(\mu)^{\dag}}
(\eta')}{E-E_{n,\mu}}
=\frac{1}{2\pi}\int_0^{2\pi}\int_0^{2\pi}
G(\varphi''\eta'',\varphi'\eta';E)
  \exp[-i\mu(\varphi''-\varphi')]d\varphi''d\varphi'. \label{eq:man3}
\ee

\par Eq.~(\ref{eq:man3}) can be viewed as
the symmetry reduced Green's function
corresponding to the azimuthal quantum number $\mu$. A semiclassical
expression for this can be obtained by using the semiclassical
approximation to the full Green's function

\begin{equation}
G(\varphi''\eta'',\varphi'\eta';E) \simeq
 \frac{2\pi}{(2\pi i\hbar)^2}\sum_{cl.tr.}
 \sqrt{\left |\det\left(D(\varphi''\eta'',\varphi'\eta';E)\right)\right |}
 \exp[\frac{i}{\hbar}S(E)-\frac{i}{2}\nu\pi] \label{eq:VV}
\end{equation}

\noindent
which has contributions from all classical trajectories ($cl. tr.$) at
energy $E$ connecting points $(\varphi'',\eta'')$ and
$(\varphi',\eta')$. The action $S(E)$ is

\bea
S(\eta'',\eta',\varphi,E)&&=
\int_{\varphi'\eta'}^{\varphi''\eta''} \bf{p.dq}\nonumber\\
    &&=\int_{\eta'}^{\eta''} p_{\eta}d\eta + p_{\varphi}(\varphi''
       - \varphi'+2\pi N)
       \nonumber\\
    &&=\overline{S}(\eta'',\eta',p_{\varphi},E)
    +p_{\varphi}(\varphi+2\pi N) \label{eq:action}
\eea

\noindent
where $p_{\varphi}$ is a constant of motion due to the axial symmetry
of  the  system,  $\varphi=\varphi''-\varphi'$,  $N$ is the winding
number and $\overline{S}$ is the reduced  action
$\int_{\eta'}^{\eta''} p_{\eta}d\eta$.
Lastly,  $\nu$ is the Maslov index which is determined
by the caustics encountered along the trajectory \cite{CLR}
while the matrix  $D$  is  given  by \cite{MCG} :

\begin{eqnarray}
D(\varphi''\eta'',\varphi'\eta';E)
&&=\frac{\partial(p_{\varphi''},p_{\eta''},t)}
        {\partial(\varphi',\eta',E)} \nonumber\cr\\
&&=\left (\matrix{{\partial^2S}
   \over{\partial\varphi'\partial\varphi''}
&{\partial^2S}\over{\partial\varphi'\partial\eta''}
&{\partial^2S}\over{\partial\varphi'\partial E}\cr
 {\partial^2S}\over{\partial\eta'\partial\varphi''}
&{\partial^2S}\over{\partial\eta'\partial\eta''}
&{\partial^2S}\over{\partial\eta'\partial E}\cr
 {\partial^2S}\over{\partial E\partial\varphi''}
&{\partial^2S}\over{\partial E\partial\eta''}
&{\partial^2S}\over{\partial E^2 }\cr}
  \right )   \label{eq:matrix}
\end{eqnarray}

\par In order to arrive at a semi-classical form for the
symmetry reduced Green's function, we insert Eq.~(\ref{eq:VV})
in Eq.~(\ref{eq:man3}) and use a new set of variables
$\varphi = \varphi '' - \varphi '$ and ${\overline \varphi}
= (\varphi '' + \varphi ')/2$ to obtain

\be
\sum_n\frac{\phi_n^{(\mu)}(\eta'')\phi_n^{(\mu)^{\dag}}
(\eta')}{E-E_{n,\mu}}  \simeq {1\over (2\pi i\hbar)^2} \sum_{cl. tr.}
\int_0^{2\pi} d\overline{\varphi}
\int_{-\pi}^{\pi} d\varphi
\exp[-i\mu\varphi]\sqrt{\left | \det(D) \right |}
\exp[{iS\over\hbar} - {i\nu\pi\over 2}]
\ee

\noindent
where we have used

\be
\int_0^{2\pi}\int_0^{2\pi}d\varphi'd\varphi''
=\int_{-\pi}^{\pi}d\varphi \int_0^{2\pi}d\overline{\varphi}.
\label{eq:rel2}
\ee

\par We shall evaluate the integral over $\varphi$ using the
method of stationary phase in the limit $\hbar \rightarrow 0$.
The stationarity condition,

\be
{\partial S\over \partial \varphi} - \mu \hbar = 0
\label{eq:select}
\ee

\noindent
restricts trajectories to quantized values of $p_\varphi$
(= $\partial S/\partial \varphi $). We shall denote the
stationary point by $\varphi^*$ such that
$p_{\varphi}(\varphi^*) = \mu\hbar$.
The integration yields

\bea
\sum_n\frac{\phi_n^{(\mu)}(\eta'')\phi_n^{(\mu)^{\dag}}
(\eta')}{E-E_{n,\mu}}  \simeq {1\over (2\pi i\hbar)^{3/2}} \sum_{cl. tr.}
\int_0^{2\pi} d\overline{\varphi} \;  &&
 {\sqrt{\left |\det\left ( D(\varphi^*,\eta'',\eta';E)\right )\right |}
\over \sqrt{\left |U \right |}} \nonumber\cr\\
&&\times  \exp[{i{\overline S}\over\hbar} - {i\nu'\pi\over 2} ]
\label{eq:phidone}
\eea

\noindent
where $U = \partial^2 S/\partial \varphi^2 =
\partial p_\varphi / \partial \varphi$ evaluated
at the stationary point $\varphi^*$ and $\nu' = \nu + \beta'$
with $\beta' = 1$ if $U$ is negative and zero otherwise.
We shall postpone the
$\overline{\varphi}$ integration but for now we merely note
that the symmetry of the system allows this to be evaluated
exactly.

\par The semi-classical symmetry reduced Green's function
is thus expressed as a sum of contributions from classical
trajectories at an energy $E$, connecting points $\eta''$
and $\eta'$ and having an angular momentum $p_\varphi = \mu \hbar$.
We can now simplify the ratio of the amplitude determinants
and express $\left |\det(D)\right | /\left | U \right |$ as  

\be
{\left |\det(D)\right |\over \left |U \right |} = 
{\left |\det(D)\right |\over \left |{\partial^2S\over 
\partial\varphi'\partial\varphi''} 
\right |} = \left | \det(\tilde{D}) \right |  \label{eq:ratio1}
\ee

\noindent
where 
\be
\tilde{D} = \left (\matrix{{\partial^2\overline{S}}
\over{\partial\eta'\partial\eta''}
& {\partial^2\overline{S}}\over{\partial\eta'\partial E}\cr
{\partial^2\overline{S}}\over{\partial E\partial\eta''} &
{\partial^2\overline{S}}\over{\partial E^2} \cr} \right )
\label{eq:tildeD}
\ee

\noindent
Details of this reduction can be found in the appendix.

\par It is useful at this point to introduce a local co-ordinate
system ($\eta_2,\eta_3$) in the reduced ($\eta$) plane  
where $\eta_2$ changes along
the trajectory while $\eta_3$ increases in the transverse
direction and is zero on the orbit. It follows from the definition 
that

\be
\frac{\partial^2\overline{S}}{\partial E \partial\eta^{''}_2}
=  \frac{1}{\dot{\eta}^{''}_2} ; \hskip 0.25 in 
\frac{\partial^2\overline{S}}{\partial\eta^{'}_2\partial E}
 =  -\frac{1}{\dot{\eta}^{'}_2 } \label{eq:man6}
\ee

\noindent
and

\be
{\partial^2 \overline{S}\over \partial\eta_i'\partial\eta_2''}  = 0 =
{\partial^2 \overline{S}\over \partial\eta_2'\partial\eta_i''}  \label{eq:man6a}
\ee

\noindent
so that the determinant ratio can be expressed as 

\be
\det(\tilde{D}) =  
\left \{ \frac{1}{\dot{\eta}^{''}_2}\frac{1}{\dot{\eta}^{'}_2}
       (-\frac{\partial^2\overline{S}}{\partial\eta^{'}_3
  \partial\eta^{''}_3})\right \}_{\varphi = \varphi^*}
=   \frac{1}{\dot{\eta}^{''}_2}\frac{1}{\dot{\eta}^{'}_2}\;R
\ee

\par With these simplifications, we are now ready to evaluate the
trace of the symmetry reduced Green's function. Setting
$\eta''=\eta'=\eta$  and integrating
over $\eta$ on both
sides of Eq.~(\ref{eq:phidone}), we get,

\bea
g_{\mu}(E) = && \sum_n\frac{1}{E-E_{n,\mu}} \nonumber\cr\\
 \simeq && {1\over (2\pi i\hbar)^{3/2}} \sum_{cl. tr.}
\int_0^{2\pi} d\overline{\varphi}\;\int d\eta \;
{1\over \dot{\eta_2}} \sqrt{\left | R \right|}
\exp[{i{\overline S}\over\hbar} - {i\nu'\pi\over 2}] 
\label{eq:red2}
\eea

\noindent
where  $g_{\mu}$  is  the  trace  of the symmetry reduced
Green's function corresponding to the azimuthal quantum number
$\mu$, $\overline{S} = \oint p_{\eta} d\eta$ and

\be
R = \left \{ - \frac{\partial^2\overline{S}}{\partial\eta^{'}_3
  \partial\eta^{''}_3}\right \}_{\varphi = \varphi^*,\eta' = \eta''}.
\ee

\par Clearly, the only orbits that contribute to the trace
are the ones that have $p_\varphi = \mu\hbar$ and
close in $\eta$ but not necessarily in
$\varphi$. Further, the $\eta$ integration can be performed
by the method of stationary phase and the stationarity
condition then picks only that subset of orbits for which
$p_{\eta'} = p_{\eta''}$ at $\eta' = \eta'' = \eta$.
We shall refer to such trajectories in the full $3D$ dynamics
as quasi-periodic trajectories (q.p.t.) since they will be
periodic in the  reduced dynamics  of  the  $\eta$-motion.
In what follows such periodic orbits in the reduced system
will  be  referred  to  as reduced periodic orbits (r.p.o.).

\par The action ${\overline S}$ in the neighbourhood of
a r.p.o. can thus be written to the second order as
$\overline{S}(\eta_3,E)  =  {\overline S}(\eta_3=0;E) +
\frac{\eta_3^2}{2} W$ where

\be
W = \Bigl[\frac
{\partial^2\overline{S}}{\partial\eta_3^{''}\partial\eta_3^{''}}
+ 2\frac
{\partial^2\overline{S}}{\partial\eta_3^{''}\partial\eta_3^{'}}
+\frac
{\partial^2\overline{S}}{\partial\eta_3^{'}\partial\eta_3^{'}}
\Bigr]_{\eta_3^{''}=\eta_3^{'}=0} \label{eq:svar}
\ee

\noindent
On performing the $\eta_3$ integration by the method of stationary
phase, Eq.~(\ref{eq:red2}) reduces to

\be
g_\mu(E)
\simeq  \frac{1}{2\pi i\hbar}\sum_{q.p.t}
\int_0^{2\pi} d\overline{\varphi} \oint {d\eta_2 \over \dot{\eta_2}}
\sqrt{\left |{R\over W} \right |_{\eta_3 = 0}}
\exp[\frac{i}{\hbar}\overline{S}(E)-i{\pi \over 2}\sigma ]
\label{eq:eta3}
\ee

\noindent
where $\sigma = \nu' + \beta$ with $\beta = 1$ if $W$ is
negative and zero otherwise. As in the case
without symmetry, $\sigma$ is the Maslov index
of the stable or unstable manifold and is an invariant of the
reduced periodic orbit \cite{CLR,Rob2}.

\par
While we do not explicitly show this, the factor involving
the second derivatives of $\overline{S}$ is related to
linearized dynamics in the neighbourhood of the r.p.o.
and is expressed as \cite{MCG}:

\begin{equation}
 \sqrt{\left | \frac{\Bigl(\frac{\partial^2\overline{S}}
 {\partial\eta^{'}_3
 \partial\eta^{''}_3}\Bigr)_{\eta^{''}_3=\eta^{'}_3=0}}
 {\Bigl(\frac
 {\partial^2\overline{S}}{\partial\eta_3^{''}\partial\eta_3^{''}}
 + 2\frac
 {\partial^2\overline{S}}{\partial\eta_3^{''}\partial\eta_3^{'}}
 +\frac
 {\partial^2\overline{S}}{\partial\eta_3^{'}\partial\eta_3^{'}}
 \Bigr)_{\eta_3^{''}=\eta_3^{'}=0}}\right |}
 =\frac{1}{\sqrt{|\det(M-I)|}}  \label{eq:linear}
\end{equation}

\noindent
where  the  $2\times  2$  matrix $M$ is the stability matrix
describing the dynamics in the linearized
neighbourhood of the r.p.o. and $I$  is  the  unit
matrix. Note that $\sigma$ is independent of the position
$\eta_2$ along the periodic orbit and so are $\overline {S}(E)$
and $\sqrt{|\det(M-I)|}$. The integration along $\eta_2$
thus yields

\begin{equation}
 \oint \frac{ d\eta_2}{\dot{\eta}_2} =T_0  \label{eq:period}
\end{equation}

\noindent
where $T_0$ is  the  period  of the primitive r.p.o., or, in terms
of the time, $T_{q.p.t.}$ required to span the
q.p.t. in the full 3D -dynamics, $T_0 = T_{q.p.t.}/N_0$
where $N_0$ is the number of primitive r.p.o. contained in
the quasi periodic trajectory. It is instructive however 
to deal with the $\eta_2,\overline{\varphi}$ integrations together
and this yields the $''$volume$''$ of initial conditions
for which the various stationarity conditions are satisfied.
Thus,

\be 
\int_0^{2\pi} d\overline{\varphi} \oint {d\eta_2
\over \dot{\eta}_2} = W_0
\ee

\noindent
where $W_0 = \overline{\Phi} T = 2\pi T_0 = 2\pi T_{q.p.t.}/N_0$.
Note that when the q.p.t. is also
periodic and has an $N_1$ fold symmetry, one can interpret
$\overline{\Phi} = 2\pi /N_1$ and $T = T_{q.p.t}/r = T_p$ 
when viewed in the full phase space. Here $r$ is the
repetition number of the periodic orbit in the
full phase space and $rN_1 = N_0$. 
With these clarifications, Eq.~(\ref{eq:eta3}) becomes,

\be
g_\mu(E) \simeq \frac{1}{2\pi i\hbar}\sum_{q.p.t}
\frac{2\pi T_0}{\sqrt{|\det(M-I)|}}
\exp{[\frac{i}{\hbar}\overline{S}(E)-\frac{i}{2}\sigma \pi]}
\label{eq:main}
\ee

\noindent
This is the main result of this section. The sum involves
quasi-periodic trajectories at an energy $E$ and with
$p_\varphi = \mu\hbar$ and is identical to the result
obtained by Creagh \cite{Creagh}.

\par It is instructive however to examine the significance
of Eq.~(\ref{eq:select}) in greater detail.
In performing the $\varphi$ integration for each orbit, it
is implicit that the action changes smoothly as the orbit
(characterized by $\varphi$ and hence $p_\varphi$) is varied.
In other words, orbits
occur in topologically equivalent families \cite{family}
within which the action varies smoothly and the
stationary phase condition of Eq.~(\ref{eq:select}) picks
out one orbit from this family. By changing
$\mu$, another orbit from the same family contributes
and this has a slightly different value
of ${\overline S}$. The range of $\mu$ over
which a family contributes depends on its extent
which in turn is decided by the potential $V(\rho,z)$.
These ideas are put on a more concrete footing
for axially symmetric cavities in the following
sections.

\par It is easy to check that the trace of the full Green's
function is recovered by summing over $\mu$ as follows :
\bea
\sum_\mu g_\mu(E) & = & \sum_N \int d\mu~\; g_\mu(E) \exp(2\pi i\mu N) \\
   &   \simeq & {1\over i\hbar} {1\over 2\pi} \sum_{N} \int d\mu
\sum_{q.p.t.}
{ 2\pi T_0 \over \sqrt{|\det(M-I)|}}
e^{{i\over \hbar}\{\overline{S} 
+ 2\pi \mu\hbar N\} - {i\sigma\pi/2}}
\eea

\noindent
where we have used the Poisson summation formula and substituted
Eq.~(\ref{eq:main}) for $g_\mu(E)$. The integral when evaluated
by the method of stationary phase picks up only those
trajectories for which
$\partial\overline{S}/\partial p_\varphi + 2\pi N = 0$ or
$\varphi^* = 2\pi N$. Thus the only orbits that contribute 
to the full spectrum are the ones which are periodic in the
full phase space.
The final expression is 

\be
g(E) \simeq \frac{1}{i\hbar(2\pi i\hbar)^{1/2}}
\sum_{p.o.}
\frac{(2\pi/N_1) T_p}{\sqrt{|\det(M-I)|}}
\sqrt{\left |\frac{\partial p_\varphi}{\partial\varphi}\right | }
\exp{[\frac{i}{\hbar}S_p(E)-\frac{i}{2}\sigma_p '\pi]}
\label{eq:CL}
\end{equation}

\noindent
with $\sigma ' = \sigma - \delta$ where $\delta = 1$ or $0$
depending on whether $\partial^2 {\overline S} /\partial p_\varphi^2$ is
positive or negative. The result is identical
to that obtained by Creagh \& Littlejohn \cite{CL1,fnote_phase}.

\par For the sake of completeness, it is necessary to mention
that there is yet another contribution to the trace that
we have so far neglected. It arises form the
{\em zero-length orbits} and gives rise to the average
density of states, $d_{av}^{\mu}(E)$ and to the leading
order this is given by \cite{LW,Creagh}~:

\be
d_{av}^\mu(E) = {1\over h^3}\int dp dq\; \hbar\;
 \delta(p_\varphi - \mu\hbar)
 \delta(E - H(p,q))  \label{eq:avden}
\ee

\noindent
Eq.~(\ref{eq:main})
gives the oscillatory part of the density,
$d_{osc}^{\mu}(E) = -{1\over \pi}\lim_{\epsilon \rightarrow 0^+}
\Im~\; g_\mu(E+i\epsilon)$ where $\Im$
denotes  the imaginary part.
In the following sections
we shall illustrate our results for axially symmetric cavities
and look at contributions of q.p.t. in $d_{osc}^\mu(E)$
as well as the $\mu$ dependence of $d_{av}^\mu(E)$.

\par Though we have restricted ourselves to the case of
axial symmetry, the formalism is very general and can
be adapted to other cases of continuous symmetry. For example
in case of rotational symmetry, one can express the wave
function as :

\be
 \psi_n^{(l,m)}(r,\theta,\varphi)= Y_l^m(\theta,\varphi)
    \phi_n^{(m,l)}(r) \label{eq:wave1}
\ee
\noindent
where $\{Y_l^m(\theta,\varphi)\}$ are the spherical
harmonics. This can be used in Eq.~(\ref{eq:bas}) to
obtain the trace of the Green's function for a given
value of the quantum numbers $l$ and $m$.

\section{Axially Symmetric  Cavity}
\label{sec:Length}

\par We now specialize to the case of a particle moving
freely inside an axially symmetric cavity and undergoing
specular reflections on collision with the wall.
As in most other systems,
it is easier to verify the duality in Eq.~(\ref{eq:main})
starting with quantum levels and obtaining the length
spectrum of periodic orbits. This is easily achieved by computing
the power spectrum of the quantum density, $d^\mu(\sqrt{E})$ :

\bea
S^\mu(l)& =& \left |\int dk d^\mu(k) \exp\{ikl\} \right |  \\
    & =& \left |\int dk (d_{av}^\mu + d_{osc}^{\mu})
       \exp\{ikl\} \right |
 \label{eq:power}
\eea

\noindent
where $\sqrt{E} = k$ and $d^\mu(k) = 2kd_\mu(k^2)$ is the
quantum density for a given value of $\mu$.
The average part of the quantum density gives rise to a
peak at zero in $S^{\mu}(l)$ while the oscillatory
part contributes peaks at the length of quasi-periodic orbits
with $p_\varphi = \mu\hbar$.

\par What is perhaps not apparent is the fact that for
$\mu \neq 0$, the reduced periodic orbits that contribute
to $d_{osc}^\mu$ vary with energy whenever they
are not periodic (i.e. $\varphi \neq 0$). These peaks
are therefore expected to shift in the power spectrum
with the window over which it is computed.

\par Consider for example an orbit in the
equatorial plane ($z=0$) where the particle reflects specularly
inside a circular billiard of radius, $R$.
It is easy to see that two successive
points on the caustic \cite{fnote2} (one reflection between them)
satisfy the conditions for quasi-periodicity since
$\rho' = \rho''$, $z' = z'' = 0$, $p_{\rho}' = p_{\rho}''$
and $p_z' = p_z'' = 0$. The action on this q.p.t
is thus~: $S_{j} = \sqrt{E} l_{j}$
where $l_{j} = 2R\sin(\varphi/2)$ where $\varphi$ is the
difference in angle, $\varphi'' - \varphi$ between these points on the
caustic. By moving away from the caustic to a point
along the trajectory, there exists another point separated by a
collision where conditions for quasi-periodicity
are satisfied (i.e. $\rho' = \rho''$ and
$p_{\rho}' = p_{\rho}''$). It is also easy to check that
along this quasi-periodic orbit, $\varphi$ is conserved
while $\varphi$ changes by moving from the caustic in the
radial direction to another orbit of the
same topological family.
The stationary phase condition in Eq.~(\ref{eq:select})
is thus :

\be
p_{\varphi} = \sqrt{E}R\cos(\varphi/2) = \mu\hbar \label{eq:select1}
\ee

\noindent
or equivalently :
\be
\varphi^* = 2\cos^{-1}(\mu\hbar/R\sqrt{E})
\ee

\noindent
so that

\be
\overline{S}_j = 2R\sqrt{E}\sqrt{1-{\mu^2\hbar^2\over R^2E}} -
2\mu\hbar\cos^{-1}(\mu\hbar/R\sqrt{E})
\ee

\noindent
in Eq.~(\ref{eq:main}). For $\mu \neq 0$, the orbit varies
with energy continuously and the system is thus inhomogeneous.
At a given (non-zero) value of $\mu$, the length of the
contributing orbit increases with energy and thus should
give rise to broadened
peaks in the power spectrum, $S^\mu(l)$.

\par Note that the stationarity condition in Eq.~(\ref{eq:select1})
picks up one orbit for each $\mu$ from this topologically
equivalent family of 1-bounce orbits within which $p_\varphi$
varies smoothly. In spite of the complicated dependence of
$\overline{S}$ on $E$ and $\mu$,
it is evident that with increasing $\mu$, peaks
in the power spectrum, $S^\mu(l)$, shift to the left.
Moreover the broadening of peaks is expected to
increase with $\mu$ due to larger spread in the length
of contributing orbits.

\par An interesting consequence of this analysis
is the fact that at almost all energies this quasi-periodic
trajectory is not eventually periodic in the full phase
space whenever $\mu \neq 0$. This follows from the fact
that periodic orbits in the circle billiard form
a set of measure zero.
This conclusion also holds in the more general
case. For $\mu \neq 0$, whenever a reduced periodic orbit
corresponds to a periodic orbit in the full phase space
with $p_\varphi = \mu\hbar$, the action, $\overline{S}_{j} =
\sqrt{E}~l_{j} - 2\pi N_{j}\mu\hbar$ where $N_{j}$ is the
winding number of the orbit. However, on changing the energy, $E$,
this orbit no longer has the quantized value of
$p_\varphi$ and hence cannot contribute. Similar arguments
hold for q.p.t's which form part of periodic orbits
in the full phase space. Thus at any energy, most
orbits that contribute to $d_{osc}^\mu$ do not
lie on a trajectory that is periodic in the full phase
space. This phenomenon is in sharp contrast to the case of
discrete symmetries where all quasi-periodic trajectories
are eventually periodic in the full phase space. The
only exception is the trivial case, $\mu = 0$
when all reduced periodic orbits lie on periodic
trajectories in the full phase space.

\par Though we have restricted ourselves
to a simple orbit in the preceding analysis,
the conclusions  are in fact more
general as the  numerical results of the following
section indicate.

\par We now return to an evaluation of the average density
of states, $d_{av}^\mu(E)$ for cavities where Eq.~(\ref{eq:avden})
becomes :

\be
d_{av}^\mu(E) = {1\over h^3}\int dp_\rho dp_z dp_\varphi
dz d\rho d\varphi \;\hbar \; \delta(p_{\varphi} - \mu\hbar)
\delta(E - p_\rho^2 - p_z^2 - {p_\varphi^2\over \rho^2})
\ee

\noindent
for a particle of mass, $M = 1/2$.
The integrals are straightforward to evaluate 
and the final
result is expressed as :

\be
d_{av}^\mu(E) = {1\over 4\pi \hbar^2} A'
\ee

\noindent
where 

\be
A' = \int_{z^\mu_{min}}^{z^\mu_{max}} dz 
\int_{\rho^\mu_{min}}^{\rho_{max}} d\rho.
\ee

\noindent
Note that $z^\mu_{min}, z^\mu_{max}, \rho^\mu_{min}$ depend on $\mu$ and
are dictated by the centrifugal barrier while $\rho_{max}$ depends
on $z$ through the shape of the cavity. For small 
$\left |\mu \right |/\sqrt{E}$ however, $A' \simeq A - Z_0 R_0$ 
where $Z_0$ is the extent
of the cavity along the $Z$-axis and $R_0 = \left |\mu \right |
\hbar/\sqrt{E}$.
Thus

\be
d_{av}^\mu(E) \simeq {A\over 4\pi}{1\over \hbar^2} - 
{\left |\mu \right | Z_0\over 4\pi} {1\over \sqrt{E}}
{1\over \hbar} \label{eq:avden1}
\ee

\noindent
where 

\be
A  = \int_{z^0_{min}}^{z^0_{max}} dz \int_0^{\rho_{max}} d\rho
\ee

\noindent 
is the area of the $\rho Z$ plane. 
Consequently, the average integrated density
of states, $N_{av}^\mu(E) = \int_0^E d_{av}^\mu(E') dE'$ is :

\be
N_{av}^\mu(E) = {AE\over 4\pi} - {\left |\mu \right |
 Z_0\over 2\pi} \sqrt{E} \label{eq:avN}
\ee

\noindent
where for convenience we have set $\hbar = 1$.

\par Note that the first term in the expression for  $N_{av}^\mu(E)$
is identical to that for
a $2-$dimensional billiard of area $A$. We have not evaluated
the standard perimeter term here which is of the same order
as the second term of Eq.~(\ref{eq:avN}), though as we shall
show later, Eq.~(\ref{eq:avN}) predicts the correct $\mu$
dependence for cavities.

\section{Some Numerical Results}
\label{sec:Numerical}

\par We present here some numerical results on an axially symmetric
cavity described by

\be
R(\theta) = {R_0\over \lambda}\{1 + \alpha_nP_n(\cos~\theta\}) 
\label{eq:cavity}
\ee

\noindent
where $R_0$ is the radius of a sphere having the same volume
as the deformed cavity, $\lambda$ is a volume preserving factor,
$\alpha_n$ is the strength of the deformation and $\{P_n\}$ are the
Legendre polynomials. 
$R(\theta)$ is the distance from the origin to a point
on the surface of the cavity, which due to the axial symmetry
depends only on the angle, $\theta$ measured from the
$Z$-axis.
For our calculations, we choose a $P_2\;(n=2)$ deformed cavity
with $R_0 = 1$, $\lambda = 1.008087$ and $\alpha_2 = 0.2$.
For convenience, we choose the mass, $M = 1/2$ and
$\hbar = 1$.
Details of the procedure used in obtaining the quantum
energy eigenvalues can be found in \cite{PM}.

\par We have made no attempt towards finding periodic
orbits systematically for this system both because there is
no obvious symbolic dynamics that one can use and also
since we are interested in the inverse problem. While
orbits in the XY plane are well known, we have
determined periodic orbits in YZ plane ($p_\varphi = 0$) using
the orbit length extremization technique. We believe, the
list we have compiled is complete for lengths, $l < 15$.
Fig.~(1) shows a plot of two periodic orbits which
also satisfy the condition $p_{\eta '} = p_{\eta ''}$ at
points separated by half their total lengths.

\par In Fig.~(2-4) we show plots of the power spectrum,
$S^{\mu}(l)$ for $\mu= (0,1),(2,4),(6,8)$. The small arrows
mark the position of full periodic orbits in the
$YZ$ plane and the orbit along the diameter in the $XY$
plane, while the larger arrows mark those orbits
that do not close in $\varphi$. All of these have
$p_{\varphi} = 0$ and hence orbits which do not close
in $\varphi$ form part of periodic orbits.
The peaks in the power spectrum for $\mu=0$ agree
very well with the lengths of quasi-periodic orbits
marked by arrows though there is no distinct peak at
$4.7615$ or $9.5230$. These correspond to the
periodic orbit along the $Z$-axis and its repetition.
Unlike other orbits which occur in a 1-parameter
family, this orbit is isolated and hence has
a much smaller contribution. We shall return
to this later while discussing states belonging to
particular parity class.

\par The $\mu=1$ power spectrum is also
described well by these orbits which have $p_\varphi = 0$.
The peaks shift to the left only slightly
indicating that the topology of the orbit does not
change while going from $p_\varphi = 0$ to a non-zero
value of $p_\varphi$. The phenomenon can be understood
from the analysis in section \ref{sec:Length} where
we have explicitly described the orbit selection
process for the simplest orbit in the $XY$ plane.
The only significant change occurs at $10.38$ where
the peak height decreases appreciably.

\par This process continues for larger values of
$\mu$ and can be seen from the power spectrum.
The peaks gradually shift to the left of the
arrow and nothing dramatic happens except that
at $\mu = 4$, there is no trace of the orbit
at $l = 10.38$. There is yet another feature
that is readily seen by comparing the power
spectra of $\mu = 0$ or $1$ and $\mu = 8$.
The peaks broaden significantly with increasing
$\mu$ even though there is hardly any
change in the range of eigenvalues considered for
evaluating the power spectrum. This is
due to the fact that the length of the orbit
(belonging to a topologically equivalent family)
varies with energy and its spread around the mean
value increases
with the value of $\mu$.

\par We emphasize this aspect in Fig.~(5) where we
plot the power spectrum for $\mu = 26$ for two
different windows of energy. The one centred at
the higher energy has peaks significantly to the right of
the other,
indicating that the lengths of contributing orbits increase
with energy.

\par We also compare the power spectrum of $\mu = 1$
and $\mu = 26$ in Fig.~(6) and find that for the simplest
(topological) orbit family (connecting successive points
on the caustic) the peaks for $\mu = 26$ are broader and
to the left of the corresponding peak for $\mu = 1$.

\par These findings corroborate our results of the
previous sections and we
now investigate the power spectrum of
even/odd parity states for a particular
value of $\mu$. Contribution to the semiclassical 
symmetry reduced even/odd
Green's function come from two sources. Apart from the usual
orbits with $p_\varphi = \mu\hbar$ connecting ($\rho',z',\varphi'$) 
and ($\rho'',z'',\varphi'')$
there are additional contributions from orbits having
$p_\varphi = \mu\hbar$ connecting points ($\rho',z',\varphi'$)
and ($\rho'',-z'',\varphi'' + \pi$) weighted
appropriately by the character of the symmetry group \cite{Rob1}.   
Alternately, one can exploit the additional $\varphi$ symmetry
and express the symmetry reduced even/odd (denoted respectively
as +/-) 
Green's function as :

\bea 
&&\sum_{n_1}\frac{\phi_{n_1}^{(\mu)}(\rho,
z)\phi_{n_1}^{(\mu)^{\dag}}(\rho, z)}{E-E_{n_1,\mu}^{\pm}}
=\frac{1}{4\pi}\Biggl[\int_0^{2\pi}\int_0^{2\pi}
G(\varphi'',\rho,                z;\varphi',\rho,                z;E)
\exp[-i\mu(\varphi''-\varphi')]d\varphi''d\varphi'\nonumber\cr\\
&&\pm (-1)^{\mu}\int_0^{2\pi}\int_0^{2\pi}           G(\varphi'',\rho,
-z;\varphi',\rho, z;E)
  \exp[-i\mu(\varphi''-\varphi')]d\varphi''d\varphi'\Biggr]
\end{eqnarray}

\noindent
The trace of its semiclassical version thus involves quasi-periodic
trajectories with $p_\varphi = \mu\hbar$ {\em and} trajectories 
connecting ($\rho,z$) with ($\rho,-z$) having $p_\varphi = \mu\hbar$,
$p_z^f = -p_z^i$ where $f$ and $i$ refer respectively to the
initial and final points.

\par
Thus  we  expect  to see both full and half q.p.t.'s in the power
spectrum  of  even  parity and odd parity sequences.
In particular, the half orbit along the $Z$-axis,
will have a lower damping (due to its instability,
the damping is exponential with the length of the
orbit) and hence could be visible in the power
spectrum. This is indeed obvious in Fig.~(7)
where a peak appears at $l=2.38$. In fact, a closer
observation also reveals a peak at $l=4.76$.

\par Significantly, orbits belonging to the XY
plane do not have any other symmetry related orbit
and hence have no additional peak at half the length.

\par Finally, we study the $\mu$ dependence of the average
density by  plotting $k_N^{(\mu)} - k_N^{(0)}$ as a function
of the level number, $N$ in Fig.~(8). Here $k_N^\mu = \sqrt{E_N^\mu}$. An
approximate expression for $k_N^{(\mu)} - k_N^{(0)}$ can be
obtained using Eq.~(\ref{eq:avN}) and is given by :

\be
k_N^{(\mu)} - k_N^{(0)} \simeq  {\left |\mu \right |
 Z_0\over A} + \sqrt{{\mu^2 Z_0^2
\over A^2} + {4\pi N\over A}} - \sqrt{ {4\pi N\over A}}
\ee

\noindent
Therefore for large $N$, $k_N^{(\mu)} - k_N^{(0)} \simeq  
{\left | \mu \right | Z_0\over A}$
and for the cavity that we have considered, this is
$1.38\left | \mu \right |$.

\par Fig.~(8) is a plot of $k_N^{(\mu)} - k_N^{(0)}$ for $\mu = 1$
(bottom curve) to $\mu = 6$ (top curve). Each curve is
approximately constant and the mean separation between
curves is $1.38 \left | \mu \right |$
as predicted.

\section{Conclusions}
\label{sec:Conclusions}

\par In the previous sections, we have studied the
semiclassics of symmetry reduced spectra for the
case of axial symmetry. There are two approaches
to the problem. The first
exploits the conservation of
$l_z$ to reduce the system to 2-dimensions.
Orbits then move around in an effective potential
that includes the centrifugal term
$\hbar^2m^2/(2M\rho^2)$. In comparison, the approach
that we adopt here uses the
dynamics in the full phase space and the
trace formula then takes a different structure
altogether. Our aim has been to understand
the nature of the classical trajectories  that
contribute and also point out the differences
with the case of discrete symmetry.

\par Most of this paper deals with a derivation
of the trace formula for the spectrum with
a given value, $\mu$ of the azimuthal quantum
number, $m$. Though this has been studied by
Creagh \cite{Creagh} earlier, we provide an independent
derivation here using only the
structure of the wavefunction to achieve
symmetry reduction.
Finally, though we have worked with axially
symmetric systems, the method can in principle
be extended to a large class of systems with
continuous symmetry.

\par Our main results  are the following :

$\bullet$ The average density, $d_{av}^ \mu (E) = 
{A\over 4\pi} - {\left |\mu \right | Z_0\over 4\pi}
{1\over \sqrt{E}}$ for small $\left | \mu \right |/\sqrt{E}$ so that
$k_N^{(\mu)} - k_N^{(0)} \simeq {Z_0\over A}$ where $N$
denotes the level number and $k_N = \sqrt{E_N}$.

$\bullet$ The only classical trajectories that
contribute to the oscillatory part of the quantum spectrum,
$d_{osc}^\mu(E)$ are those which close in
$(\rho,z,p_\rho,p_z)$ and for which $p_\varphi = \mu\hbar$.
Thus they need not be closed in the full phase
space though they may be eventually periodic when
allowed to evolve. These orbits are however
periodic in the reduced phase space $(\rho,z,p_\rho,p_z)$.

$\bullet$ For $\mu = 0$, all reduced periodic orbits
are identical to or part of periodic orbits in the
full phase space.

$\bullet$ For $\mu \neq 0$, almost all orbits that
contribute to $d^\mu(E)$ at any given energy
are non-periodic in the full phase space even
when they are allowed to evolve in time. Peaks in the
power spectrum thus shift with the window
over which it is computed and the width of peaks
increases with $\mu$.

\par Our numerical results support these observations
and allow us to state the following :

$\bullet$ for smooth potentials (billiards with
smooth boundaries), topologically equivalent orbits
contribute at successive values of $\mu$
as can be seen from the gradual shift in the
position of peaks in the power spectrum.
This is implicit in the selection
criterion in Eq.~(\ref{eq:select}) since each of
the orbits selected necessarily occurs in
a (topological) family within which $p_\varphi$
varies smoothly.

\par The transition from $\mu = 0$ to $\mu \neq 0$
is therefore smooth and this might have some significance
in quantizing the spectrum for $\mu = 1$ using
information about contributing orbits at $\mu = 0$.

\section{Acknowledgements}

D.B acknowledges several useful and interesting
discussions with Bertrand
Georgeot and Gregor Tanner and thanks Stephen
Creagh for several clarifications.

\appendix
\section{}

\par We outline here the steps involved in the reduction
of the amplitude determinants leading to Eq.~(\ref{eq:ratio1}).

\par The equation for the reduced action

\be
\overline{S}(\eta'',\eta',p_{\varphi},E) = S(\eta'',\eta',\varphi,E)
- p_{\varphi} \varphi  \label{eq:A1}
\ee

\noindent
is a Legendre transformation from the variables $(\eta'',\eta',\varphi,E)$
to the variables $(\eta'',\eta',p_{\varphi},E)$ and leads to the relations

\be
{\partial S \over \partial\eta'_i} = {\partial \overline{S}\over
\partial\eta'_i} ; \hskip 0.25 in
{\partial S \over \partial\eta''_i} = 
{\partial \overline{S}\over \partial\eta''_i} ; \hskip 0.25 in
{\partial S\over \partial E} = {\partial \overline{S} \over \partial E} ;
\hskip 0.25 in
{\partial S\over \partial \varphi} = p_\varphi ;  \hskip 0.25 in
{\partial\overline{S} \over \partial p_\varphi} = -\varphi . \label{eq:Legen}
\ee

\par The following expressions involving the second 
derivatives of the action can now be derived :

\begin{mathletters}
\be
{\partial^2S\over \partial\eta'_i\partial\eta''_j} = 
{\partial^2\overline {S}\over \partial\eta'_i\partial\eta''_j} +
{\partial^2\overline {S}\over \partial p_\varphi\partial\eta'_i}
{\partial^2\overline {S}\over \partial p_\varphi\partial\eta''_j}
{\partial^2S\over \partial \varphi^2} \label{eq:rela} 
\ee
\be
{\partial^2S\over \partial\eta'_i\partial E} = 
{\partial^2\overline {S}\over \partial\eta'_i\partial E} +
{\partial^2\overline {S}\over \partial p_\varphi\partial\eta'_i}
{\partial^2\overline {S}\over \partial p_\varphi\partial E}
{\partial^2S\over \partial \varphi^2}  \label{eq:relb}
\ee
\be
{\partial^2S\over \partial\eta''_i\partial E} = 
{\partial^2\overline {S}\over \partial\eta''_i\partial E} +
{\partial^2\overline {S}\over \partial p_\varphi\partial\eta''_i}
{\partial^2\overline {S}\over \partial p_\varphi\partial E}
{\partial^2S\over \partial \varphi^2}  \label{eq:relc}
\ee

\be
{\partial^2S\over \partial E^2}  = 
{\partial^2\overline {S}\over \partial E^2} +
{\partial^2\overline {S}\over \partial p_\varphi\partial E}
{\partial^2\overline {S}\over \partial p_\varphi\partial E}
{\partial^2S\over \partial \varphi^2} \label{eq:reld}
\ee
\be
{\partial^2S\over \partial \varphi^2} = - \left ({\partial^2\overline{S}
\over \partial p_\varphi^2} \right )^{-1}  \label{eq:rele}
\ee
\be
{\partial^2S\over \partial\eta'_i\partial\varphi} = 
{\partial^2\overline {S}\over \partial p_\varphi\partial\eta'_i}
{\partial^2S\over \partial \varphi^2} \label{eq:relf}
\ee
\be
{\partial^2S\over \partial\eta''_i\partial\varphi} = 
{\partial^2\overline {S}\over \partial p_\varphi\partial\eta''_i}
{\partial^2S\over \partial \varphi^2} \label{eq:relg}
\ee
\be
{\partial^2S\over \partial\varphi\partial E} = 
{\partial^2\overline {S}\over \partial p_\varphi\partial E}
{\partial^2S\over \partial \varphi^2} \label{eq:relh}
\ee
\end{mathletters}

\noindent
We shall derive Eq.~(\ref{eq:rela}) as an illustration.

\bea
{\partial^2S\over \partial\eta'_i\partial\eta''_j} && =  
{\partial\over \partial\eta'_i}\left({\partial S\over 
\partial \eta''_j}\right) \nonumber \cr 
&& \cr
&& = {\partial\over \partial\eta'_i}\left({\partial\overline{
S} \over \partial\eta''_j}(\eta'',\eta',p_\varphi,E) \right) \nonumber \cr
&& = {\partial^2\overline{S}\over\partial\eta'_i\partial\eta''_j} +
{\partial^2\overline{S}\over\partial p_\varphi\partial\eta''_j}
{\partial p_\varphi \over \partial\eta'_i} \label{eq:stp1}
\eea

\noindent
Also,

\bea
{\partial^2\overline{S} \over \partial p_\varphi \partial\eta'_i} && =
{\partial \over \partial p_\varphi}\left ({\partial S
\over \partial\eta'_i}(\eta'',\eta',\varphi,E) \right ) \nonumber \cr
&& \cr
&& = {\partial ^2S\over\partial\varphi\partial\eta'_i}
{\partial \varphi\over\partial p_\varphi} \nonumber
\eea

\noindent
so that
\be
{\partial p_\varphi \over \partial\eta'_i} = {\partial^2\overline{S}
\over\partial p_\varphi\partial\eta'_i}
{\partial^2S\over\partial\varphi^2}.   \label{eq:stp2}
\ee

\noindent
Thus,
\be
{\partial^2S\over \partial\eta'_i\partial\eta''_j} = 
{\partial^2\overline {S}\over \partial\eta'_i\partial\eta''_j} +
{\partial^2\overline {S}\over \partial p_\varphi\partial\eta''_j}  
{\partial^2\overline {S}\over \partial p_\varphi\partial\eta'_i}
{\partial^2S\over \partial \varphi^2} \label{eq:stp3}
\ee

\par We are now in a position to deal with the ratio of determinants.
By interchanging rows and columns, the determinant 
of Eq.~(\ref{eq:matrix}) can be expressed as 

\be
\det(D)  = \det \left (\matrix{{\partial^2S}
   \over{\partial\eta'\partial\eta''}
&{\partial^2S}\over{\partial\eta'\partial E}
&{\partial^2S}\over{\partial\eta'\partial \varphi''}\cr
 {\partial^2S}\over{\partial E\partial\eta''}
&{\partial^2S}\over{\partial E^2}
&{\partial^2S}\over{\partial E\partial \varphi''}\cr
 {\partial^2S}\over{\partial \varphi'\partial\eta''}
&{\partial^2S}\over{\partial \varphi'\partial E}
&{\partial^2S}\over{\partial \varphi'\partial \varphi'' }\cr}
  \right )
= {\partial^2S\over \partial\varphi'\partial\varphi''}
\det(\overline{D} - \overline{E}) \label{eq:expand1}
\ee

\noindent
where 

\be
\overline{D} = \left ( \matrix{
{\partial^2S\over \partial\eta'\partial\eta''} &
{\partial^2S\over \partial\eta'\partial E } \cr
{\partial^2S\over \partial E\partial\eta'} &
{\partial^2S\over \partial E^2}}\right )  \label{eq:mat1}
\ee

\noindent
and

\bea
\overline{E} && = \left ( \matrix{
{\partial^2S\over \partial\eta'\partial\varphi''} \cr
{\partial^2S\over\partial E\partial\varphi''}}\right )
\left ( {\partial^2S\over\partial\varphi'\partial\varphi''}\right )^{-1}
\left (\matrix{ {\partial^2S\over\partial\varphi'\partial\eta''} &
{\partial^2S\over\partial\varphi'\partial E}}\right ) \nonumber \cr 
&& \cr
&& = \left ( {\partial^2S\over\partial\varphi'\partial\varphi''}\right )^{-1}
\left ( \matrix{
{\partial^2S\over\partial\eta'\partial\varphi''}
{\partial^2S\over\partial\varphi'\partial\eta''} &
{\partial^2S\over\partial\eta'\partial\varphi''}
{\partial^2S\over\partial\varphi'\partial E} \cr
{\partial^2S\over\partial E\partial\varphi''}
{\partial^2S\over\partial\varphi'\partial\eta''} &
{\partial^2S\over\partial E\partial\varphi''}
{\partial^2S\over\partial\varphi'\partial E}}\right ) \label{eq:mat2}
\eea

\noindent
Using relations (\ref{eq:rela})-(\ref{eq:relh}), it follows that

\be
\det(\tilde{D}) = {\det{D}\over 
{\partial^2S\over\partial\varphi'\partial\varphi''}} =  
\det(\overline{D} - \overline{E}) = 
\det \left ( \matrix{
{\partial^2\overline{S}\over\partial\eta'\partial\eta''} &
{\partial^2\overline{S}\over\partial \eta'\partial E} \cr
{\partial^2\overline{S}\over\partial E\partial\eta''} &
{\partial^2\overline{S}\over\partial E^2}\cr} \right ) \label{eq:final}
\ee

\pagebreak

\begin{figure}
\end{figure}
\begin{figure}
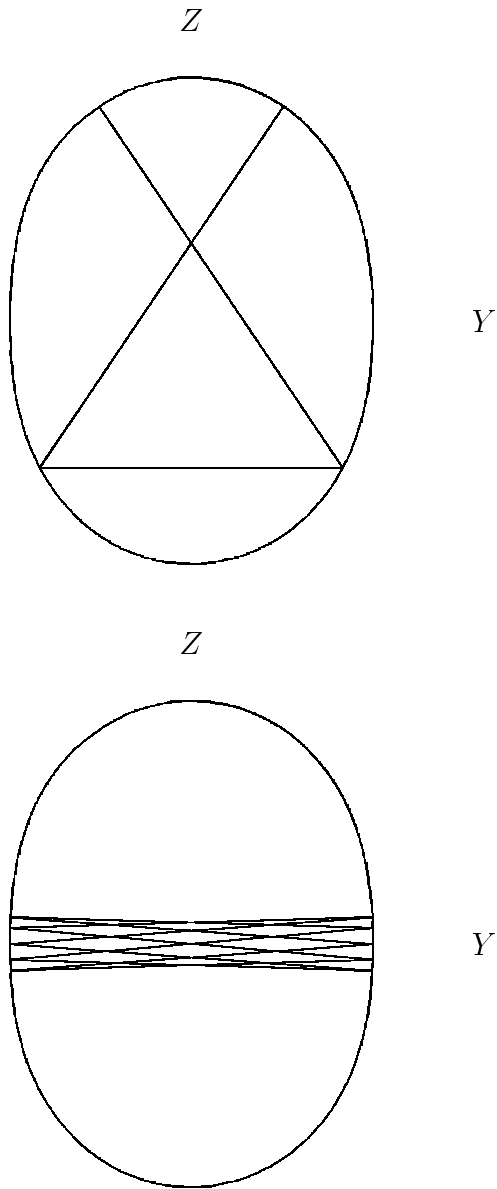

\caption{Two periodic orbits in the $YZ$ plane on which 
points separated by half the total length
satisfy the conditions, $z'=z'',\rho'=\rho''$
and $p_{z'} = p_{z''}, p_{\rho'} = p_{\rho''}$.
Their respective lengths are 11.532715
and 17.856224.}
\end{figure}
\vskip 0.15 in
\begin{figure}
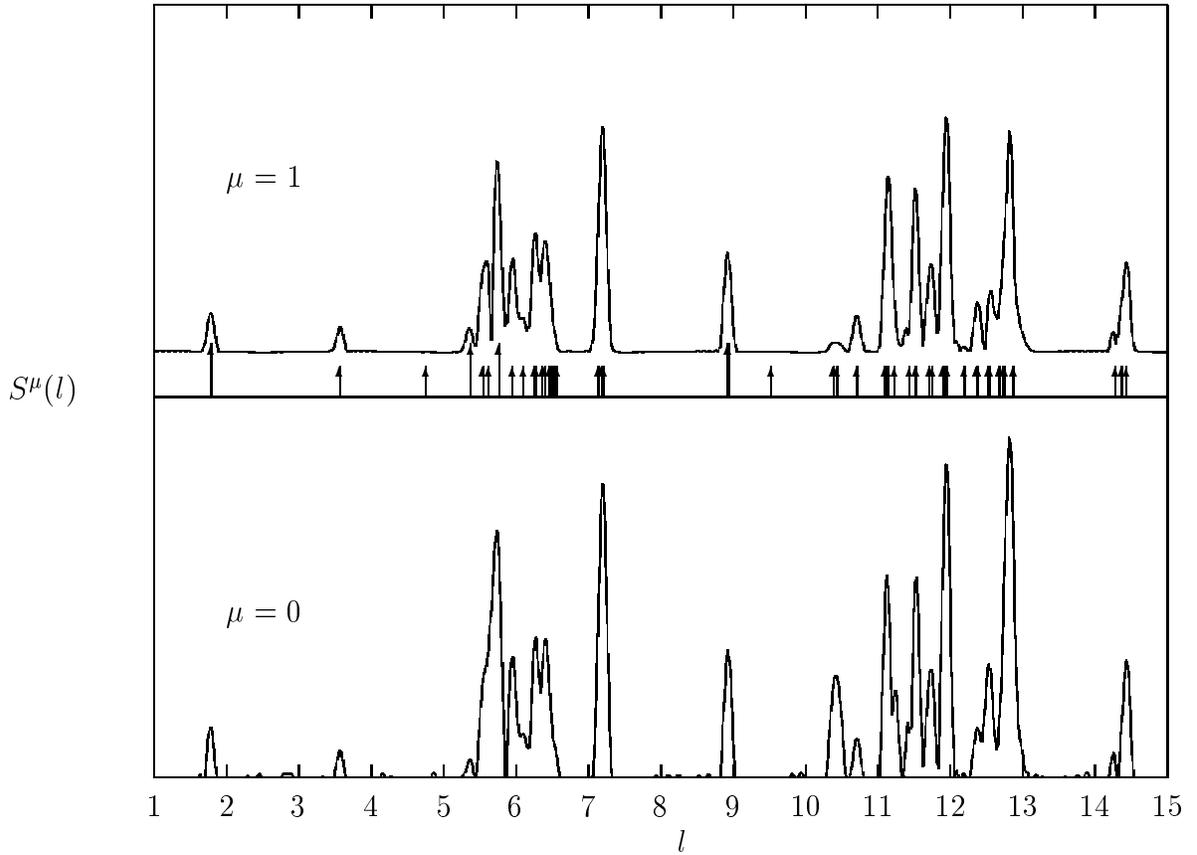

\caption{Power spectrum, $S^\mu(l)$ of $\mu=0 \; \&\; 1$ states.
The shorter set of arrows mark the lengths of full periodic
orbits in the $YZ$ plane while the longer arrows are quasi-periodic
trajectories in the $YZ$ plane that do not close in $\varphi$.
All trajectories have $p_\varphi = 0$.}  
\end{figure}
\vskip 0.15 in
\begin{figure}
\caption{Power spectrum, $S^{\mu}(l)$ of $\mu=2 \; \& \; 4$ states.
The arrows mark the length of orbits with $p_\varphi = 0$ as in
Fig.~(2). }
\end{figure}
\vskip 0.15 in
\begin{figure}
\caption{Power spectrum, $S^\mu (l)$ of $\mu= 6 \; \& \; 8$ states.
The arrows mark the length of orbits with $p_\varphi = 0$ as in
Fig.~(2). }
\end{figure}
\vskip 0.15 in
\begin{figure}
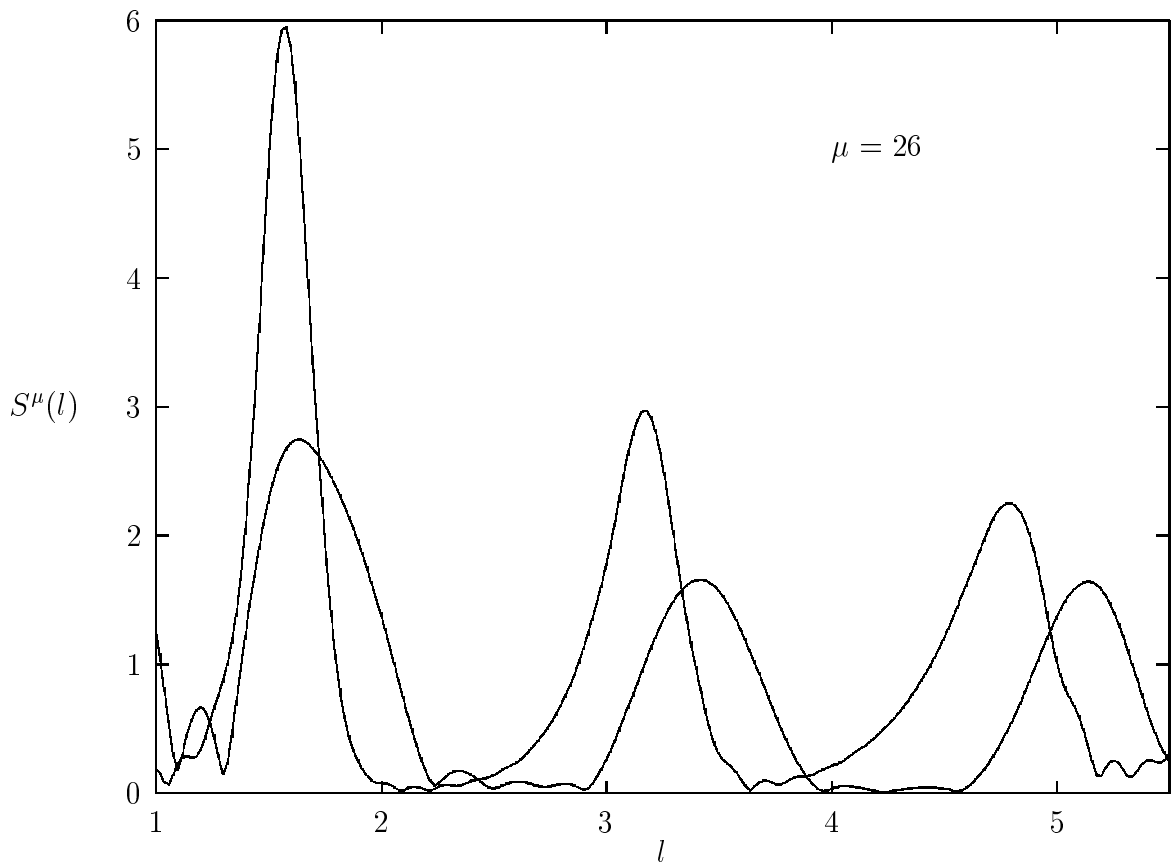

\caption{Power spectrum of $\mu=26$~states for two different 
ranges of energy. The window centred at higher energy has peaks 
shifted to the
right.}
\end{figure}
\vskip 0.15 in
\begin{figure}
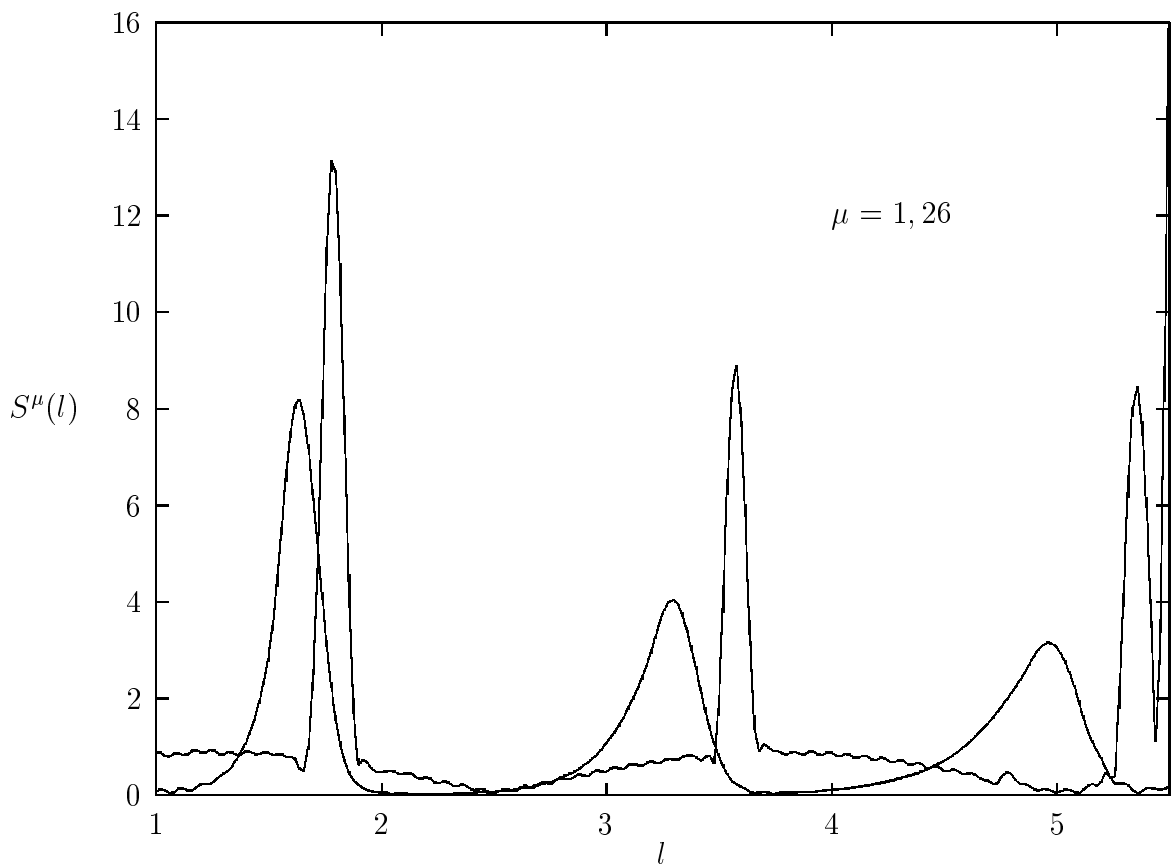

\caption{A comparison of the power spectrum of $\mu=1$ and $26$ states.
The $\mu=26$ spectrum gives rise to broader peaks due to a larger spread
in the lengths of contributing orbits. The window is
approximately over the same range of energy in both cases.}
\end{figure}
\vskip 0.15 in
\begin{figure}
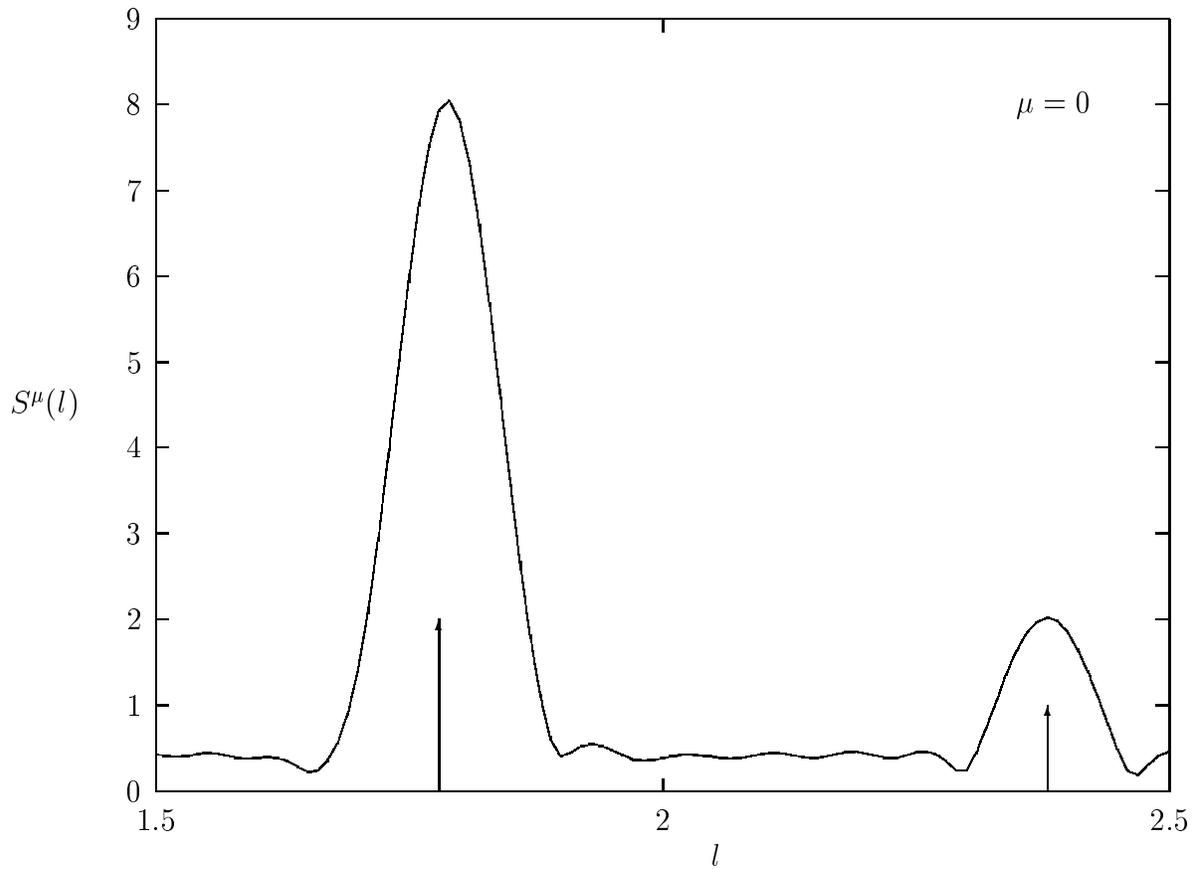

\caption{Power spectrum of $\mu=0 $, even parity states. The smaller
arrow marks the length of the half-orbit along the $Z$-axis.}
\end{figure}
\vskip 0.15 in
\begin{figure}
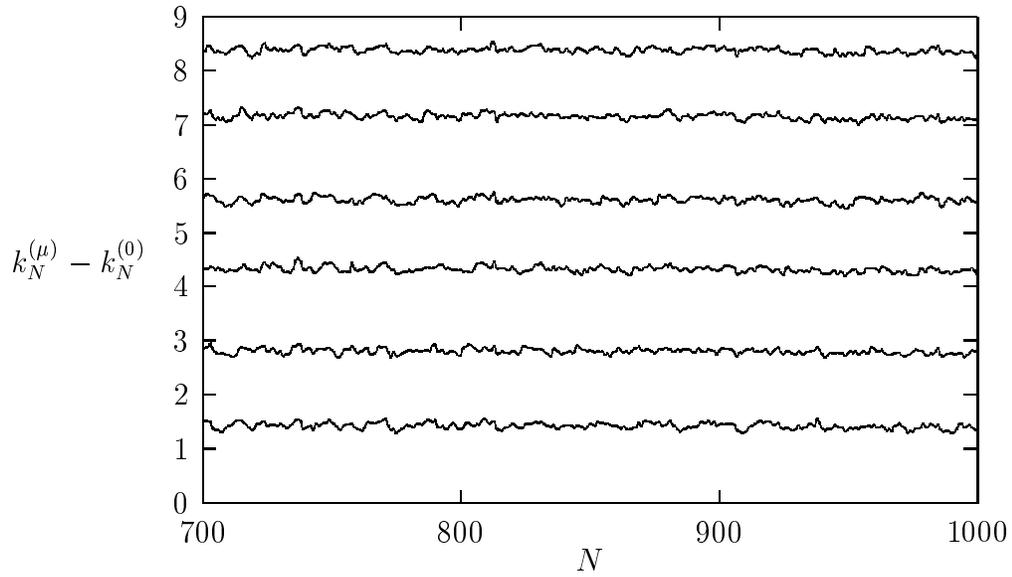

\caption{A plot of $k_N^{(\mu)} - k_N^{(0)}$ as a function
of the level number, $N$. The curves from bottom to top are
(in order of increasing $\mu$) for $\mu = 1$ to $\mu = 6$.} 
\end{figure}


\begin{references}
\bibitem{MCG} M.C.Gutzwiller, {\it Chaos in Classical and Quantum Mechanics},
Springer, New York, 1990~; in {\it Chaos and Quantum Physics},
Les Houches 1989, eds. M.-J.Giannoni, A.Voros and J.Zinn-Justin,
North Holland, 1991.
\bibitem{Rob2} J.M.Robbins, Nonlinearity, {\bf 4}, 343(1991); S.Creagh,
J.M.Robbins and R.G.Littlejohn, Phys. Rev. {\bf A42}, 1907(1990).
\bibitem{CL1} S.Creagh and R.Littlejohn,
Phys. Rev. {\bf A44}, 836(1991)
\bibitem{CL2} S.Creagh and R.Littlejohn, J. Phys. {\it A}: Math \& Gen
{\bf 25}, 1643(1992).
\bibitem{PJ} P.J.Richens and M.V.Berry, Physica D{\bf 2}, 495(1981).
\bibitem{Rob1} J.M. Robbins, Phys.Rev. {\bf A40}, 2128(1989).
\bibitem{CE} P.Cvitanovi\'{c} and B.Eckhardt, Nonlinearity {\bf 6}, 277
(1993).
\bibitem{cylindrical} $(\rho,\varphi,z)$ are cylindrical co-ordinates.
In the corresponding Cartesian coordinate system,
the axis of symmetry is the $Z$ axis
and the $z=0$ plane is the $XY$ plane.
\bibitem{WF} H.Friedrich and D.Wintgen, Phys.Rep. {\bf183}, 37(1989).
\bibitem{Creagh} S.Creagh, J.Phys. {\bf A26}, 95(1993).
\bibitem{CLR} S.C.Creagh, J.M.Robbins and R.G.Littlejohn, Phys. Rev. {\bf
A 42}, 1907(1990).
\bibitem{family} The phrase {\it topologically equivalent family}
is used in a sense that is distinct from the  usual {\it 1-parameter
family} in which all orbits exist due to the axial symmetry and
within which the action is constant.
\bibitem{fnote_phase} See also 
page 65 of 
S.Creagh, Ann. Phys. {\bf 248}, 60 (1996). The apparent difference
between the phase appearing therein and the present phase $\sigma '$
arises due to different ways of defining the azimuthal angle but
nevertheless they lead to the same value for both the phases. 
\bibitem{LW} B.Lauritzen and N.Whelan, Ann. Phys. (N.Y), {\bf 244},
112(1995).
\bibitem{fnote2} For a circle, this is the point of nearest approach
from the centre.
\bibitem{PM} T.Mukhopadhay and S.Pal, Nucl. Phys. {\bf A592}, 291(1995).


\end{references}
\end{document}